\documentclass[a4paper]{jpconf}

\usepackage[natbib=true, backend=biber, style=numeric, doi=false, isbn=false, url=false, sorting=none, firstinits=true]{biblatex}

\DeclareNameAlias{sortname}{last-first}
\DeclareNameAlias{default}{last-first}

\usepackage{fullwidth}

\usepackage[utf8]{inputenc}
\usepackage[T1]{fontenc}
\usepackage[english, french]{babel} 
\usepackage{amsmath}
\usepackage{amsfonts}
\usepackage{makeidx}
\usepackage{graphicx}
\usepackage{mathtools} 
\usepackage{braket} 
\usepackage[linkcolor=black, citecolor=black]{hyperref} 
\usepackage[usenames,dvipsnames]{xcolor} 
\usepackage{caption} 

\usepackage{tikz} 
\usetikzlibrary{positioning}
\usepackage{ dsfont } 

\newcommand{\gv}{\ensuremath{\mathbf{g}}}
\newcommand{\sub}{\ensuremath{M}}

\newcommand{\id}{\ensuremath{\text{idos}}}

\begin{document}

\title{Gap structure of 1D cut and project Hamiltonians.}
\author{Nicolas Macé, Anuradha Jagannathan and Frédéric Piéchon}
\address{Laboratoire de Physique des Solides, Université Paris-Saclay, 91400 Orsay, France}
\ead{nicolas.mace@u-psud.fr}

\selectlanguage{english}

\date{\today}

\begin{abstract}
We study the gap properties of nearest neighbors tight binding models on quasiperiodic chains.
We argue that two kind of gaps should be distinguished: stable and transient.
We show that stable gaps have a well defined quasiperiodic limit. 
We also show that there is a direct relation between the gap size and the gap label.
\end{abstract}

\section*{Introduction}

The gap labeling theorem (GLT) \cite{bellissard} is one of the few exact results known about electronic properties of quasicrystals.
It concerns the \emph{integrated density of states} (IDOS), which is defined as $\id(E) = \text{fraction of energy states below energy~}E$.
Inside a spectral gap, the IDOS is of course constant, and the theorem constrains the value it can take.
In the case of canonical cut and project chains, the theorem states that
\begin{equation}
\label{eq:gl}
	\id(E \in \text{gap}) = \frac{n}{1+\alpha} \mod 1 = \frac{n}{1+\alpha} + m
\end{equation}
where $\alpha$ is the irrational involved in the construction of the quasiperiodic chain. 
$m$ and $n$ are integers, $n$ is the label of the gap, and $m(n)$ is such that the IDOS satisfies $0 \leq \id \leq  1$.

The GLT can appear mysterious and profound. However, it has a simple derivation, making use of the approximants.
Let $\alpha_l = p_l/q_l$ be a convergent of $\alpha$. 
Consider the periodic approximant built with $\alpha_l$.
Its spectrum consists in general of $N_l = p_l + q_l$ energy bands \cite{diffractionLuck, codimSire}, and thus the IDOS inside a gap is given by
\begin{equation}
	\id(E \in \text{gap}) = \frac{j(E)}{N_l}
\end{equation}
where $j(E)$ counts the number of bands below energy $E$.
Since $p_l$ and $q_l$ are relatively prime, using Bézout's identity we can write $j = a p_l + b q_l = (a-b) q_l + b N_l$.
We let $n = a-b$ and we remark that
\begin{equation}
\label{eq:glapprox}
	\id(E \in \text{gap}) = n \frac{q_l}{N_l} \mod 1
\end{equation}
Taking the quasiperiodic limit $l \to \infty$ at fixed $n$, we recover the GLT.
Moreover, the formula \eqref{eq:glapprox} extends the GLT to approximants, which, as we will see, provides useful insight into the gap structure.

As such, the theorem contains little information: it only constrains the set of values the IDOS can take inside the gaps, but does not tell anything about the energy location of the gaps, or about their width.
Nevertheless, as we shall see, the gap label $n$ contains additional information the GLT does not explicitly tell us about.

\section{Nearest neighbors tight-binding model on quasiperiodic chains}

We consider an electron hopping on a quasiperiodic chain, obtained by the canonical cut and project method \cite{baakegrimm}.
As a generic example, we consider a model where the nearest neighbors hopping amplitude takes two values according to the quasiperiodic sequence:
\begin{equation}
	H(\alpha) = \sum_{x \in \mathbf{Z}} t_{x,x+1} \ket{x} \bra{x+1} + \text{h.c.}
\end{equation}
where $t_{x,x+1} = t_1$ or $t_{x,x+1} = t_2$.
Once the irrational is fixed, the only free parameter in the model is the ratio of the two jump amplitudes $\rho = t_1 / t_2$.

This model can be realized experimentally with light \cite{light} or cavity polaritons \cite{polaritons}.
In the latter case, the details of the energy spectrum could be accurately measured, leading the authors to observe subtle phenomena such as the energy displacement of an edge state within a spectral gap.
Such observation makes theoretical studies of the spectral gaps particularly relevant.

\section{Application: the Fibonacci quasicrystal}

\begin{figure}[htp]
	\centering
	\begin{minipage}{0.4\textwidth}
		\centering
		\includegraphics[width=1.\textwidth]{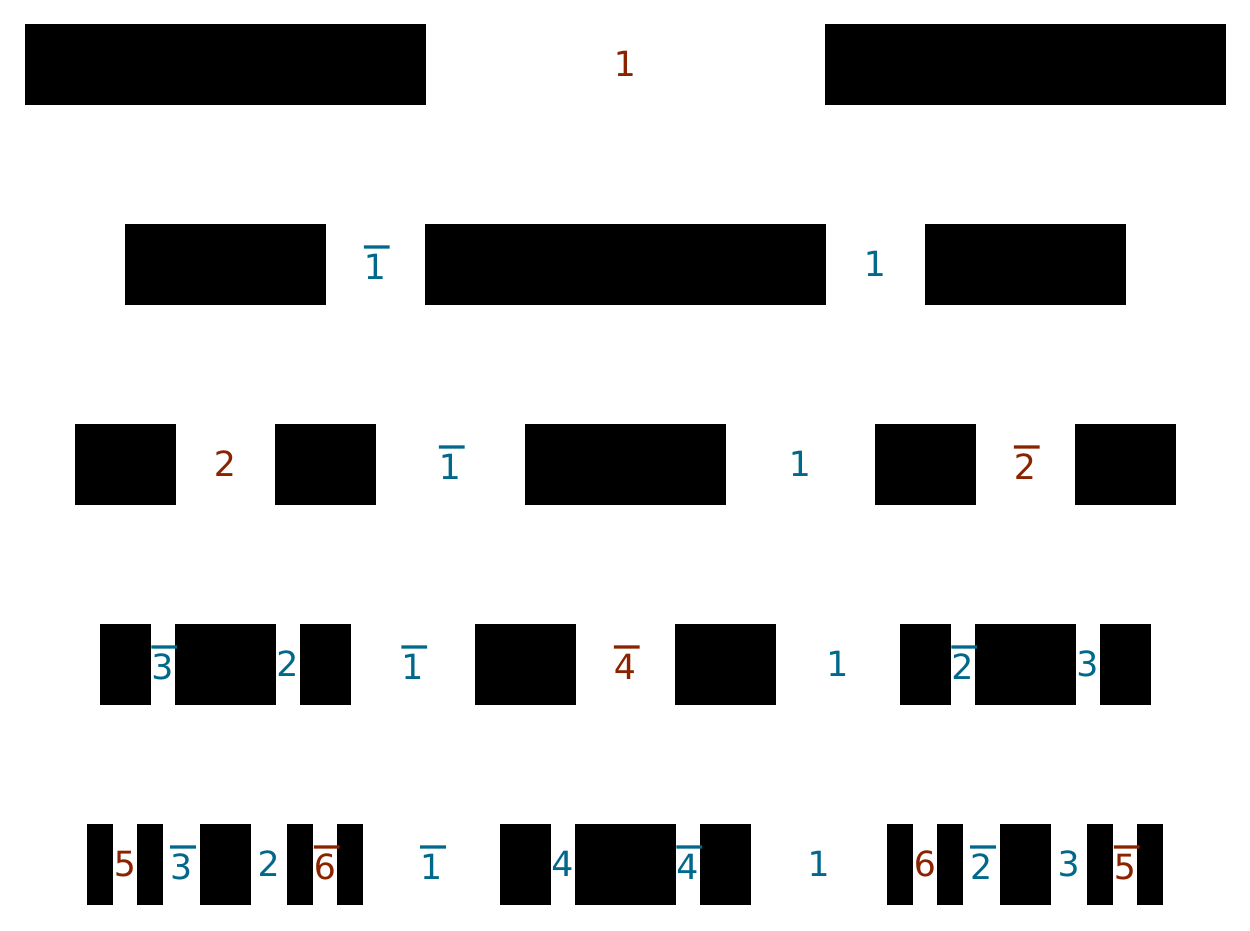}
		\caption{\small {The gap labels of the first few Fibonacci approximants. Blue: stable gaps, red: transient gaps.}}
		\label{fig:fibogaps}
	\end{minipage}~~
	\begin{minipage}{0.4\textwidth}
		\centering
		\includegraphics[width=1.\textwidth]{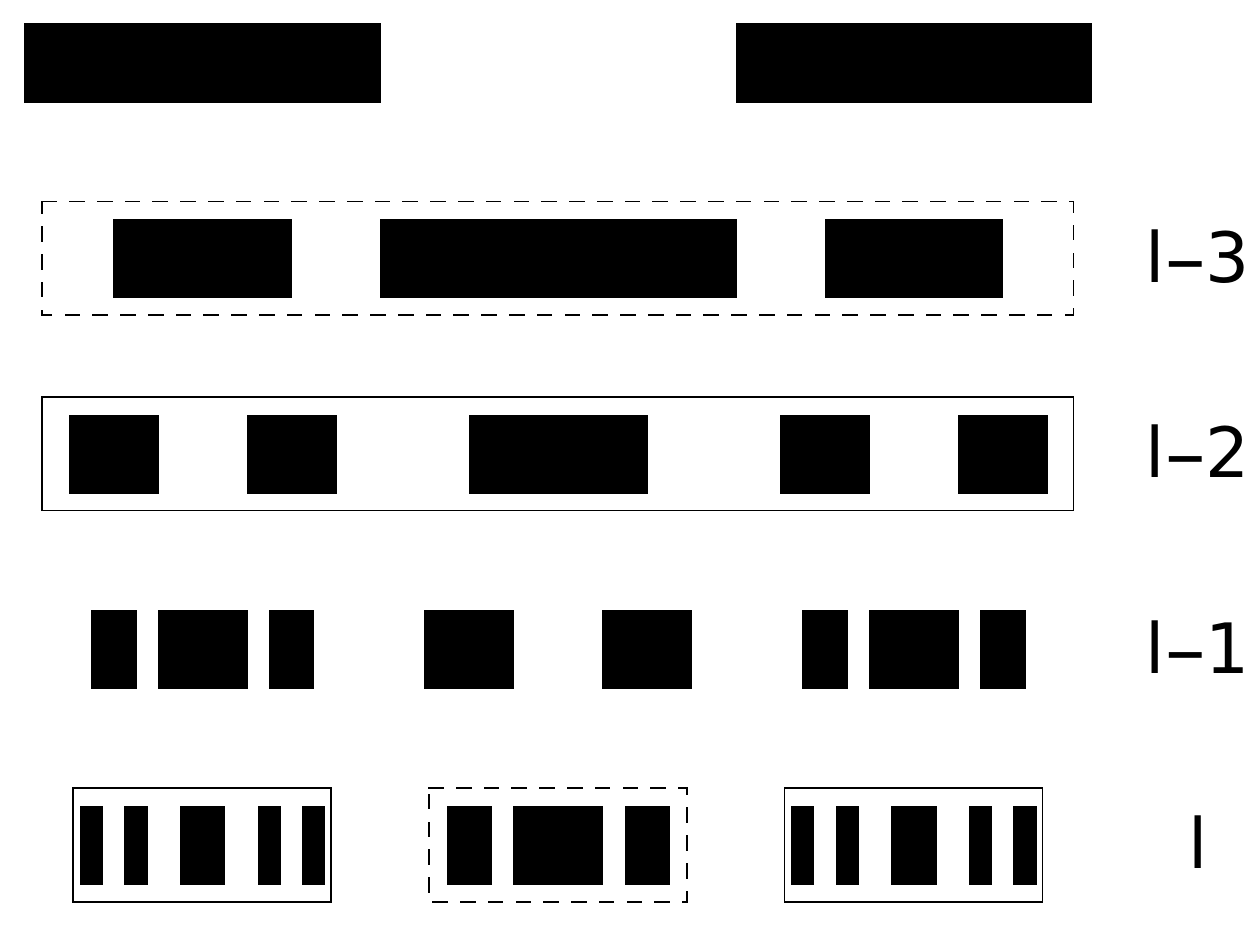}
		\caption{\small{The a recursive construction of the Fibonacci spectrum.}}
		\label{fig:fibospec}
	\end{minipage}
	\center{\small{Energy spectra of the first few Fibonacci approximants. Here we took $\rho = 0.5$, and the spectra were approximated using the renormalization group method \cite{Niu90, Piechon95}.}}
\end{figure}

We now examine in detail the special case of the Fibonacci chain, which will give us more insights.
It is characterized by $\alpha = \omega$, where $\omega = (\sqrt{5}-1)/2$ is the inverse of the golden ratio.
The gap labeling is thus
\begin{equation}
	\label{eq:fibogaps}
	\id(E \in \text{gap}) = n \omega \mod 1
\end{equation}
The approximants are built with consecutive Fibonacci numbers: $p_l = F_{l-2}$, $q_l = F_{l-1}$.
The $l^\text{th}$ approximant has $N_l-1 = F_l -1$ gaps.
Naively, we could expect all these gaps to remain open in the quasiperiodic limit $l \to \infty$.
This is not true. 
For example, the approximants of the form $l = 3m$ have a gap at $E=0$ (fig \eqref{fig:fibogaps}), but it closes in the quasiperiodic limit, as can be seen directly on the approximants, or by the fact that there is no $n$ such that $n \omega \mod 1 = 1/2$.
We will say that such a gap is \emph{transient}.
On the contrary, the two largest gaps of the Fibonacci approximants (labeled $n = \pm 1$ see fig \eqref{fig:fibogaps}) appear to be \emph{stable}: they stay open all the way to the quasiperiodic limit.

The behavior of an electron of the Fibonacci chain is well understood. 
One can show \cite{Niu90, Piechon95} that the spectrum of the $l^\text{th}$ approximant can be approximated by a simple linear combination of the spectra of the approximants $l-2$ and $l-3$ (fig \eqref{fig:fibospec}).
This provides us with a recursive gap labeling procedure. 
Let $\gv = (m, n)$ be the vector containing the labels of a gap \eqref{eq:gl}.
Let $G_l$ be the set of all gap labels $\gv$ of the $l^\text{th}$ approximant.
Using the recursive spectrum construction of figure \eqref{fig:fibospec} we see that $G_l$ can be partitioned into 3 subsets containing the gap labels of the left, central and right clusters of energy bands.
We have the following recursive relations \cite{fiboperturbMace}:
\begin{align}
	G_{l}^{\text{left}} &= \sub^{-2} G_{l-2} \\
	G_{l}^0 &= \sub^{-3} G_{l-3} + \gv_1\\
	G_{l}^\text{right} &= \sub^{-2} G_{l-2} + \gv_2
\end{align}
where $\gv_1 = (1, -1)$ and $\gv_2 = (0, 1)$ are the labels of the two main gaps (corresponding to $n = \pm 1$, see figure \eqref{fig:fibogaps}),
and $\sub$ is the geometrical substitution matrix
\begin{equation}
	\sub = 
	\begin{bmatrix}
		1 & 1\\
		1 & 0\\
	\end{bmatrix}
\end{equation}
that can be used to generate the Fibonacci sequence.

This recursive relation provides us with a very simple geometrical interpretation of the GLT, in the simple case of the Fibonacci chain.
Moreover, the construction shows that the transient gaps are the iterates of the $E=0$ gap, while the stable gaps are the iterates of the two main gaps.

This example shows that the GLT can be extended to the approximants, at the price of introducing transient gaps that are not present in the spectrum of the quasiperiodic system. 
Fortunately, we are able to pinpoint them.
We now turn to the general characterization of the transient and stable gaps in the case of an arbitrary cut and project chain.
\section{Transient and stable gaps}

\begin{figure}[htp]
	\centering
	\begin{minipage}{0.3\textwidth}
		\centering
		\includegraphics[width=1.\textwidth]{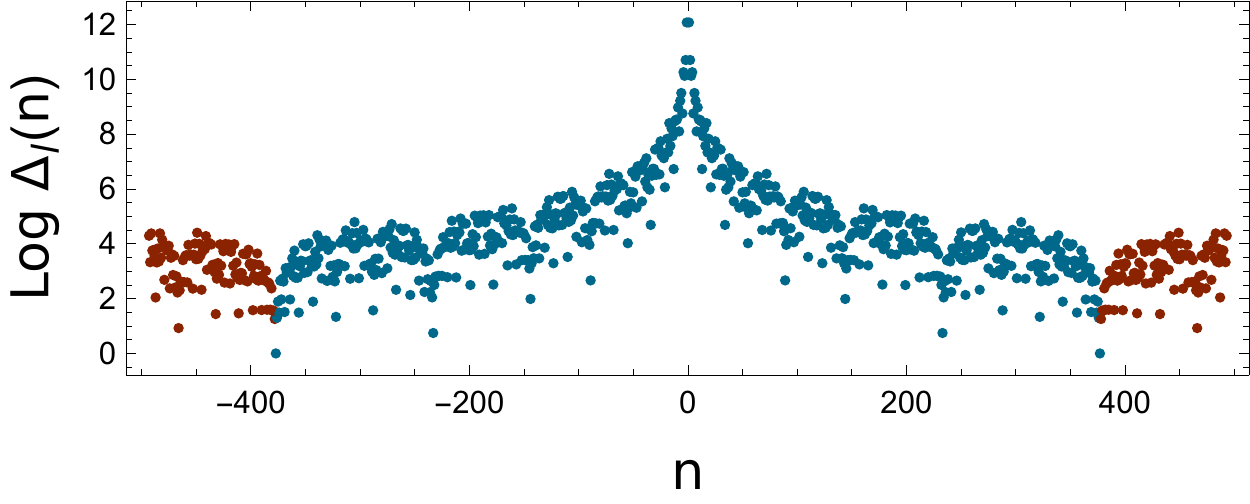}
		\caption{\small{Fibonacci chain of $N_{16} = 987$ atoms.}}
		\label{fig:width1}
	\end{minipage}~~
	\begin{minipage}{0.3\textwidth}
		\centering
		\includegraphics[width=1.\textwidth]{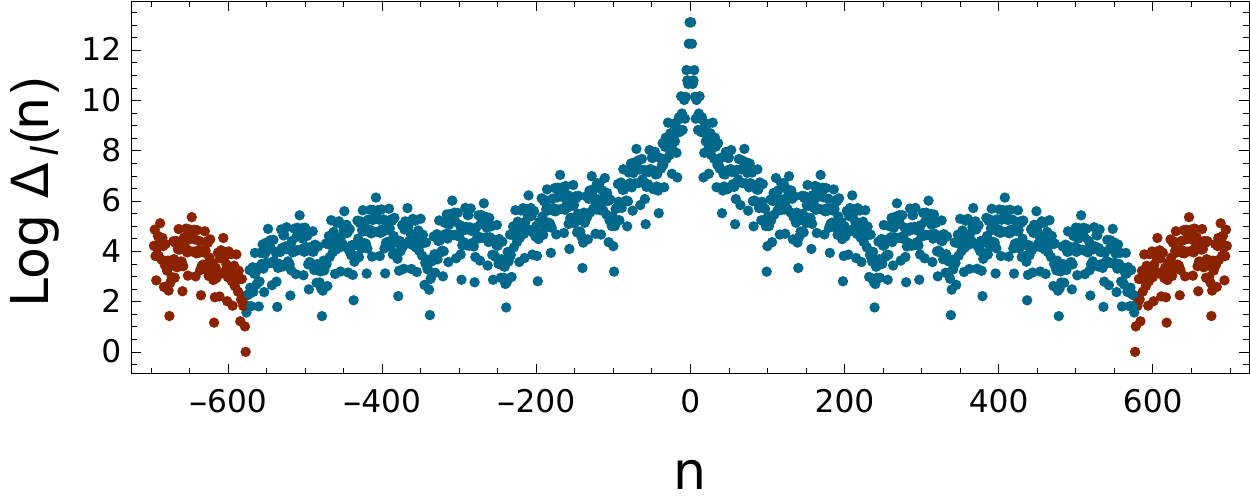}
		\caption{\small{Silver mean chain of $N_{10} = 1393$ atoms.}}
		\label{fig:width2}
	\end{minipage}~~
	\begin{minipage}{0.3\textwidth}
		\centering
		\includegraphics[width=1.\textwidth]{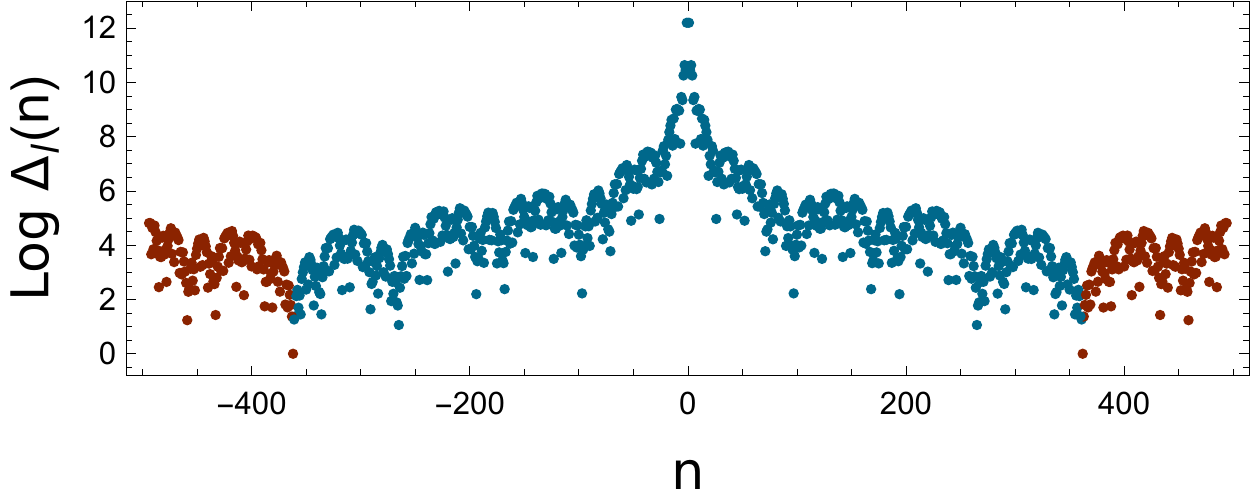}
		\caption{\small{Chain $\alpha = \sqrt{3}-1$ of $N_{12} = 989$ atoms.}}
		\label{fig:width3}
	\end{minipage}
	\center{\small{Gap width as a function of gap label for several quasiperiodic chains. Blue: stable gap, red: transient gaps, according to criterion \eqref{eq:overlapcriterion}. The ratio of the couplings was set arbitrarily to $\rho = 0.5$.}}
\end{figure}

In the previous part, we considered a gap centered around energy $\overline{E}$ to be transient if its width $\Delta_l(\overline{E})$ vanished:
\begin{equation}
\label{eq:limitcriterion}
	\Delta_l(\overline{E}) \xrightarrow[l \to \infty]{} 0
\end{equation}
We introduce here a more practical criterion.
Let $\overline{E}_l(n)$ be the mean energy inside the gap of label $n$, for the $l^\text{th}$ approximant.
Define the displacement of a gap as $D_l(n) = |\overline{E}_{l+1}(n) - \overline{E}_l(n)| $
We say the gap $n$ at step $l$ is transient if its half-width does not overlap significantly its counterpart at step $l+1$, ie if 
\begin{equation}
\label{eq:overlapcriterion}
	a(\rho) ( \Delta_l(n) + \Delta_{l+1}(n) ) < D_l(n),
\end{equation}
and stable otherwise.
We have introduced the \emph{overlap coefficient} $a(\rho)$, which quantify how much overlap we require for a gap to be considered stable.
Numerically, we determine $a$ by requiring that any stable gap at step $l$ overlaps their counterpart at step $l+L$ with $L$ arbitrarily large.
As the ratio of the couplings $\rho$ is decreased, gaps are enlarged and overlap more and more.
Thus, the value of $a$ also decreases with $\rho$ (we thank the referee for pointing out that fact).
As an example, we find $a(0.5) = 0.25$ for the Fibonacci chain.

It is easy to see that criterion \eqref{eq:overlapcriterion}, applied to the Fibonacci chain, yields the same stable and transient gaps as criterion \eqref{eq:limitcriterion} 
It is not trivial that these two criteria are in general equivalent.
However, it does appear to be the case for all the quasiperiodic chains we tested numerically. 
We will therefore consider them equivalent from now.

Figures \eqref{fig:width1} \eqref{fig:width2} \eqref{fig:width3} show the widths of the gaps as a function of their label, for various quasiperiodic chains.
The global structure is similar in every case: the gap width is a globally decreasing function of the gap index (this is consistent with perturbation theory in the limit $\rho \to 1$, see eg. \cite{Luck}).
Here we also observe that width decreases as a power-law of the index.
In the case of the Fibonacci chain, we notice log-periodic oscillations, stemming from the scale invariance of the Fibonacci Hamiltonian.
In all cases, there is a critical label above which all gaps are transient, and below which all gaps are stable.


\section*{Conclusion and perspectives}

We have shown that the well known GLT is applicable not just to quasiperiodic chains but also to their approximants.
The price to pay is the introduction of \emph{transient gaps}, which vanish in the quasiperiodic limit. 
Fortunately, these gaps are easy to spot: they always have the highest gap label.
Moreover, we show that the gap label contains relevant physical information: it globally orders the gaps by decreasing width, and separate stable from transient gaps.

The gap label, which had \emph{a priori} no physical content, thus appears physically relevant.
It has also been shown the gap label can be considered as a topological number, and used to characterize edge states \cite{Levy2015}.

Open problems include more finely describing the behavior of the gap width with the index and extending this description to higher dimensional quasicrystals, which are also known to have gaps \cite{Zijlstra, penrosegaps}.

\begin{fullwidth}
\bibhang=50pt
\printbibliography
\end{fullwidth}

\end{document}